\documentclass[twocolumn,pra, showpacs, noshowkeys,nofootinbib]{revtex4}

\usepackage{latexsym}
\usepackage{amssymb}
\usepackage{graphicx}
\usepackage{amsmath}
\usepackage{amsfonts}
\usepackage[english]{babel}
\usepackage{epsfig}
\usepackage{subfigure}
\usepackage[T1]{fontenc}
\usepackage{graphics}
\usepackage{graphicx, lscape}
\usepackage{supertabular}
\usepackage{eucal,amsfonts}
\usepackage{amsbsy}
\usepackage{psfrag}

\usepackage{color}

\def\E[#1]{{\rm E}\left[#1\right]}
\def\Ec[#1|#2]{{\rm E}\left[#1|#2\right]}

\def\lb{\lambda}
\def\M(#1){\mathbb{#1}}

\def\P[#1]{{\rm P}\left[\,#1\,\right]}
\def\Pc[#1|#2]{{\rm P}\left[#1|#2\right]}

\usepackage{
amssymb,
}

\def\b(#1){\langle#1|}
\def\B(#1){{\bf #1}}
\def\C(#1){{\cal #1}}

\def\E[#1]{{\rm E}\left[#1\right]}
\def\Ec[#1|#2]{{\rm E}\left[#1|#2\right]}

\def\ket(#1){|#1\rangle}
\def\bk(#1,#2){\langle#1|#2\rangle}
\def\bok(#1,#2,#3){\langle#1|#2|#3\rangle}
\def\kb(#1,#2){|#1\rangle
\langle#2|}
\def\lb{\lambda}
\def\M(#1){\mathbb{#1}}

\def\P[#1]{{\rm P}\left[\,#1\,\right]}
\def\Pc[#1|#2]{{\rm P}\left[\,#1\,|\,#2\,\right]}

\begin{document}

\thispagestyle{empty}

\title{Revisiting the Dolinar Receiver through\\[1mm]
Multiple--Copy State Discrimination Theory}

\date{\today}

\author{Antonio \surname{Assalini}}
\author{Nicola \surname{Dalla Pozza}}
\author{Gianfranco \surname{Pierobon}\vspace{2mm}
}

\affiliation{Department of Information Engineering (DEI), University
of Padova, \\ Via Gradenigo 6/B, 35131, Padova, Italy\\ 
\textit{e}-{\rm mail:$\,$\textit{name.surname}@dei.unipd.it}}


\begin{abstract}
We consider the problem of discriminating between two quantum coherent states by interpreting a single state like being a collection of several successive copies of weaker coherent states. By means of recent results on multiple-copy state discrimination, it is possible to give a reinterpretation of the Dolinar receiver, and carry out a quite straightforward analysis of its behavior. We also propose and investigate a suboptimal detection scheme derived from the Dolinar's architecture, which is shown to slightly outperform some other near-optimal schemes available in literature.
\end{abstract}




\maketitle


\section{Introduction}\label{Sect:1}
Discrimination between two non--orthogonal quantum states is a
fundamental issue in quantum mechanics and, in particular, in quantum communications. From a theoretical point of view, the problem was completely analyzed and solved by Helstrom \cite{Helstrom}, which found the optimal measurement operators, and the corresponding correct detection probability (Helstrom bound), for both pure and
mixed quantum states. Unfortunately, also for pure states, often the optimal measurements do not correspond to  quantum observables that are easily measurable, so that the experimental implementation of the optimal discrimination is a very difficult task.

For the case of two coherent states of a traveling single radiation mode, in 1973 Dolinar \cite{Dolinar} proposed an adaptive measurement scheme, based on a combination of photon counting and feedback control, that precisely achieves the Helstrom bound (see also \cite{Holevo}). However, since the scheme requires a very precise control of an optical--electrical loop, only recently the Dolinar's idea has obtained a satisfactory practical implementation \cite{Cook}.

Recent years have seen an increasing interest for adaptive
measurements from both a theoretical and an experimental
point of view, also for optical phase measurements and estimation
(see \cite{Wiseman} and references therein). A notably interesting
theoretical result has been obtained by Acin {\it et al.}
\cite{Acin} for discrimination between pure quantum states, when
multiple identical copies of a quantum state are available. They proved (see
also \cite{Brody}) that in this case the Helstrom bound can be achieved
by local adaptive measurements applied to single copies.
The result is particularly attractive in that it offers a
useful insight into the Dolinar's approach for discrimination between
coherent states. In this paper we discuss the strict connection,
already recognized in \cite{Wiseman}, between the ideas underlying
measurements of multiple copies of pure states \cite{Acin} and the
Dolinar receiver \cite{Dolinar}.

Acin {\it et al.} \cite{Acin} considered the discrimination between
pure quantum states $\ket(\gamma_0)$ and $\ket(\gamma_1)$, when $n$
identical copies of an unknown state are given. Formally,
the problem consists in discriminating between the pure
states
\begin{equation}
\begin{split}
\ket(\alpha_0)=\ket(\gamma_0)\otimes\cdots\otimes\ket(\gamma_0)\\ \ket(\alpha_1)=\ket(\gamma_1)\otimes\cdots\otimes\ket(\gamma_1)
\end{split}
\label{eq:1}
\end{equation}
in the tensorial product Hilbert space  $\C(H)^{\otimes n}$,
where $\C(H)$ is the Hilbert space spanned by the single
copies $\ket(\gamma_0)$ and $\ket(\gamma_1)$. Of course, also
in this case one could apply the Helstrom theory and find
an optimal collective  measurement in $\C(H)^{\otimes n}$
achieving  the Helstrom bound. Unfortunately, collective
measurements are difficult to realize experimentally. With
the adaptive approach suggested in \cite{Acin}, the experimenter
performs on each copy a local measurement which is optimized
on the basis of the results of the measurements on the
previous copies. The surprising enough conclusion is that
the optimized local measurements achieve the Helstrom bound,
exactly the same as the optimal collective measurement.

The discrimination between two coherent states of a
single--mode harmonic oscillator (without
loss of generality $\ket(\gamma)$ and ${\ket(-\gamma)}$)
presents a difficulty similar
to that of collective measurements on multiple copies. Namely,
the Helstrom theory gives optimum measurement vectors that are
linear superposition of $\ket(\gamma)$ and ${\ket(-\gamma)}$ and do not
correspond to any measurable observable. On the other hand,
owing to their peculiar properties, the coherent states $\ket(\gamma)$
and ${\ket(-\gamma)}$ of duration $T$ can be thought as sequences of
shorter and weaker modes of duration $T/n$, namely,
\begin{equation}
\begin{split}
\ket(\gamma)=&\left|\frac{\gamma}{\sqrt n}\right\rangle
\otimes\cdots\otimes\left|\frac{\gamma}{\sqrt n}\right\rangle\\[2mm]
{\ket(-\gamma)}=&\left|-\frac{\gamma}{\sqrt n}\right\rangle
\otimes\cdots\otimes\left|-\frac{\gamma}{\sqrt n}\right\rangle\;.
\end{split}
\label{eq:2}
\end{equation}
Moreover, as $n$ increases, and the average number of photons
per copy goes to zero, the optimal Helstrom measurement on
each copy may be conveniently approximated by a displacement
followed by a photon detection. Then, in principle, we may
think to apply  the multiple--copy adaptive measurement to
the segmented quantum states \eqref{eq:2}. As $n$ goes to infinity,
it appears natural the transition to the Dolinar scheme \cite{Dolinar},
with a continuous time--varying displacement controlled by
the photon counting results.

This paper is organized as follows. In Section~\ref{Sect:2} we
illustrate the optimal multiple--copy measurement paradigm from
a novel point of view leading in a natural way to
the continuous--time extension as in the Dolinar receiver.
In Section~\ref{Sect:3} the feasibility of near--optimal discrimination
of weak coherent states is discussed. In Section~\ref{Sect:4}
we extend the optimal multiple--copy adaptive measurements to
coherent states. In particular, we derive the theory of the
Dolinar receiver in a simple way, avoiding the cumbersome
machinery of dynamic programming. Finally, in Section~\ref{Sect:5} we propose a suboptimal simplified version of the Dolinar's scheme.


\section{Multiple Copy Adaptive Measurement}\label{Sect:2}
Measurement strategies for discrimination between multiple
copies of two pure quantum states were discussed
by Acin {\it et al.} \cite{Acin} (see also \cite{Brody}). Alice, accordingly
to the binary input symbol $a\in\{0,1\}$, chooses between two pure
states $\ket(\gamma_0)$ and $\ket(\gamma_1)$ in the Hilbert space
$\C(H)$ with probability $q_0$ and $q_1=1-q_0$, respectively.
(Without loss of generality we assume $q_0\ge q_1$). Then,
Alice sends Bob $n$ identical copies of the chosen state,
corresponding to the states \eqref{eq:1}, which are pure states in
the tensorial product Hilbert space $\C(H)^{\otimes n}$. Bob
performs a measurement on the system, trying to guess the
original state with maximum correct detection probability.

The optimal result is given by the well--known Helstrom bound
\cite{Helstrom}
\begin{equation}
\begin{split}
P_c&=\frac{1}{2}\left[1+\sqrt{1-4q_0q_1X^2}\right]\\&=\frac{1}{2} \left[1+\sqrt{1-4q_0q_1\chi^{2n}}\right]
\end{split}
\label{eq:3}
\end{equation}
where
\begin{equation}
X=|\bk(\alpha_0,\alpha_1)|=|\bk(\gamma_0,\gamma_1)|^n=\chi^n\label{eq:4}
\end{equation}
and $\chi=|\bk(\gamma_0,\gamma_1)|$ is the overlap coefficient of
the single copies. Bob may achieve this bound by a global
measurement using suitable von Neumann projectors $\Pi_0=
\kb(\beta_0,\beta_0)$ and $\Pi_1=\kb(\beta_1, \beta_1)$ over
the product space $\C(H)^{\otimes n}$. Unfortunately, the
optimum measurement vectors $\ket(\beta_0)$ and $\ket(\beta_1)$
turn out to be non separable linear superposition of the
pure states $\ket(\alpha_0)$ and $\ket(\alpha_1)$, i.e., an entangling
measurement which is hard to implement experimentally.

The problem of optimizing local adaptive measurements had been
tackled by Acin {\it et al.} \cite{Acin} and it may be formalized in the
following way. Let us  assume, without loss of generality, that
the single copies are given by
\begin{equation}
\begin{split}
\ket(\gamma_0)&=\cos\theta\ket(x)+\sin\theta\ket(y)\\
\ket(\gamma_1)&=\cos\theta\ket(x)-\sin\theta\ket(y)
\end{split}\label{eq:5}
\end{equation}
where $\ket(x)$ and $\ket(y)$ form an orthonormal basis of the
Hilbert space $\C(H)$ spanned by the states $\ket(\gamma_0)$
and $\ket(\gamma_1)$ and the overlap coefficient is given by
\begin{equation}
\chi=\bk(\gamma_0,\gamma_1)=\cos2\theta\;.\label{eq:6}
\end{equation}
Assume that the local measurement orthonormal vectors
on the $k$-th copy, $k=1,\ldots,n$,
\begin{equation}
\begin{split}
\ket(\mu_{k0})&=\cos\phi_k\ket(x)+\sin\phi_k\ket(y)\\
\ket(\mu_{k1})&=\sin\phi_k\ket(x)-\cos\phi_k\ket(y)
\end{split}\label{eq:7}
\end{equation}
are completely specified by the measurement angle $\phi_k$. Let $z_k
\in\{0,1\}$ be the outcome of the $k$-th measurement, which is also assumed
as the result of a provisional decision. The adaptive optimization problem
consists in finding a starting measurement angle $\phi_1$ and a recursive rule
\begin{equation}
\phi_k=f_k(z_1,\ldots,z_{k-1})\label{eq:8}
\end{equation}
in such a way that the final outcome $z_n$ gives the correct detection
with maximum probability. In the recursion \eqref{eq:8} the information gained
by the previous $k-1$ measurements is exploited in order to optimize the
choice of the next measurement angle.

This appears to be a dynamic programming problem \cite{Bertsekas} and, as
such, it had been dealt with and solved in \cite{Acin}. The main
results, surprisingly simple, are summarized as follows: i) the optimal solution of the dynamic programming approach gives
correct detection probability coinciding with the Helstrom bound \eqref{eq:3},
so that the global optimal measurement may be replaced by more
easily implementable local measurements; ii) the problem reduces to a bayesian updating problem (see also \cite{Brody}) with recursive relation
$
\phi_k=f_k(z_{k-1})
$
so that the new optimal measurement angle depends only on the
outcome of the last measurement; iii) if the $(k-1)$--th result is $z_{k-1}=i$, the optimal measurement angle $\phi_k$ is the solution of the Helstrom
optimization problem obtained replacing the a priori
probabilities $q_0$ and $q_1$ with the a posteriori
probabilities $\Pc[a=0|z_{k-1}=i]$ and $\Pc[a=1|z_{k-1}=i]$, respectively.

In particular, after the measurement on the $(k-1)$--th copy, the
provisional correct detection probability coincides with the
Helstrom bound on $k-1$ copies, namely,
\begin{equation}
P_c^{(k-1)}=\P[z_{k-1}=a]=\frac{1}{2}\left[1+\sqrt{1-4q_0q_1\chi^{2(k-1)}}\right]\;.
\label{eq:9}
\end{equation}
The next measurement angle $\phi_k$ for the $k$--th copy is chosen
to maximize the probability of correct detection under the assumption that
the a priori probabilities $q_0$ and $q_1$ are replaced by the corresponding
a posteriori probabilities of the input simbol, given the last result $z_{k-1}$.
These turn out to be $\P[a=i|z_{k-1}=i]=P_c^{(k-1)}$, $i=0,1$. Finally,
the measurement angles are given by
\begin{equation}
\begin{split}
\phi_k=\frac{1}{2}\arctan\left[\frac{1}{\sqrt{1-4q_0q_1\chi^{2(k-1)}}}
\tan2\theta\right]\quad,\\ \hspace{3cm} k=1,\ldots,n
\end{split}
\label{eq:10}
\end{equation}
if $z_{k-1}=0$ and $\pi/2-\phi_k$ if $z_{k-1}=1$. A simple proof of these
results is given in the Appendix.\footnote{Note
that the above results cannot be extended to multiple
copies of mixed states \cite{Higgins}.}

The optimum local adaptive measurement can be summarized by the following
step--by--step procedure.

\begin{itemize}
\item[1.] From the overlap coefficient $\chi$ and the input
probabilities $q_0$ and $q_1$ compute the two sequences
of measurement angles
\begin{equation}
\begin{array}{cccc}
  \hspace{8mm}\phi_1 \hspace{8mm}&\hspace{8mm} \phi_2 \hspace{8mm}& \ldots &\hspace{8mm} \phi_n\hspace{8mm} \\
  \pi/2-\phi_1 & \pi_2-\phi_2 & \ldots & \pi/2-\phi_n
\end{array}\label{eq:11}
\end{equation}
\item[2.] Start with angle $\phi_1$ if $q_0\ge1/2$ (and $\pi/2
-\phi_1$ otherwise).

\item[3.] Use the angles of the first sequence \eqref{eq:11} as long as the measurement result is 0.

\item[4.] Change angle sequence every time the result changes and
accept $z_n$ as the global result.

\end{itemize}

We will show in the sequel that this paradigm is mimicked in a continuous
time version by the Dolinar receiver.

As a further comment, we note that the multiple--copy optimization
requires the discrimination between two hypotheses in the
$2^n$--dimensional Hilbert space $\C(H)^{\otimes n}$, so that
the optimal solution is not uniquely defined. On the contrary,
the Helstrom solution discriminates between the hypotheses in
the restricted subspace $\C(H)_0$ spanned by $\ket(\alpha_0)$
and $\ket(\alpha_1)$. Of course, each optimal measurement in
$\C(H)^{\otimes n}$, once projected in $\C(H)_0$, returns the
Helstrom projectors. In particular, in the above procedure,
to each sequence $z_1,\ldots,z_n$ of results it corresponds a
measurement vector $\ket(\mu_1)\otimes\ldots\otimes\ket(\mu_n)$
in $\C(H)^{\otimes n}$ with measurement angles chosen in the
sequences \eqref{eq:11}. It can be easily verified that the $2^n$
measurement vectors in $\C(H)^{\otimes n}$ are orthonormal
and they globally give a von Neumann projective measure.


\section{Coherent Single--Copy Measurement}\label{Sect:3}
Now we consider the possibility of applying the multiple--copy adaptive
approach to the discrimination between two coherent states segmented
like in \eqref{eq:2}. Let us suppose that Alice prepares a single copy of binary coherent
states. Without loss of generality, we can assume $\ket(\alpha_0)=\ket(\gamma)$
and ${\ket(\alpha_1)=\ket(-\gamma)}$, with $\gamma$ real, as in the Binary Phase Shift Keying (BPSK)
modulation scheme, with overlap coefficient
\begin{equation}
X=|\bk(\alpha_0,\alpha_1)|=e^{-2\gamma^2}\;,\label{eq:12}
\end{equation}
where $\gamma^2$ represents the average number of photons in each state.
A straightforward application of the Helstrom theory leads to the Helstrom bound
\begin{equation}
P_c=\frac{1}{2}\left[1+\sqrt{1-4q_0q_1e^{-4\gamma^2}}\right]\;,\label{eq:13}
\end{equation}
but the corresponding measurement vectors are difficult to implement. Then,
several near--optimal, but simpler to implement, detection schemes had been devised
in the past.

A well--known solution is given by the  Kennedy receiver \cite{Kennedy}. Bob applies
a displacement $D(-\gamma)$ to the Alice state and tests the resulting state
with a photon counter. The displaced states become $\ket(\alpha'_0)=\ket(0)$
and ${\ket(\alpha'_1) =\ket(-2\gamma)}$. The photon counting detection corresponds
to measurement projectors $\Pi_0=\kb(0,0)$ and $\Pi_1=I-\Pi_0$. Decision
$a=1$ is accepted if the photon counter clicks, otherwise $a=0$ is chosen. The correct detection probability reads
\begin{equation}
\begin{split}
P_K=&q_0\bok(\alpha'_0,\Pi_0,\alpha'_0)+q_1\bok(\alpha'_1,\Pi_1,\alpha'_1)\\=&
q_0+q_1(1-e^{-4\gamma^2})\;.
\end{split}
\label{eq:14}
\end{equation}

An improved version \cite{Wittmann} of the Kennedy receiver employs
a displacement $D(-\beta)$ to be optimized, so that
the correct detection probability becomes
\begin{equation}
\begin{split}
P_{IK}=&q_0\bok(\gamma-\beta,\Pi_0,\gamma-\beta)+
q_1\bok(\gamma+\beta,\Pi_1,\gamma+\beta)\\=&q_0 e^{-(\gamma-\beta)^2}
+q_1\left(1-e^{-(\gamma+\beta)^2}\right)\;.
\end{split}
\label{eq:15}
\end{equation}
By nulling the derivative with respect to $\beta$, we
find that the displacement quantity  $\beta_0$ maximizing
\eqref{eq:15} satisfies the transcendental equation
\begin{equation}
\frac{q_0}{q_1}=\frac{\beta_0+\gamma}{\beta_0-\gamma}\, e^{-4\beta_0\gamma}\;,\label{eq:16}
\end{equation}
that can be numerically solved, and the corresponding
value of $P_{IK}$ evaluated.

\begin{figure}[t]
\centering
     \epsfxsize 0.9\hsize\epsfbox{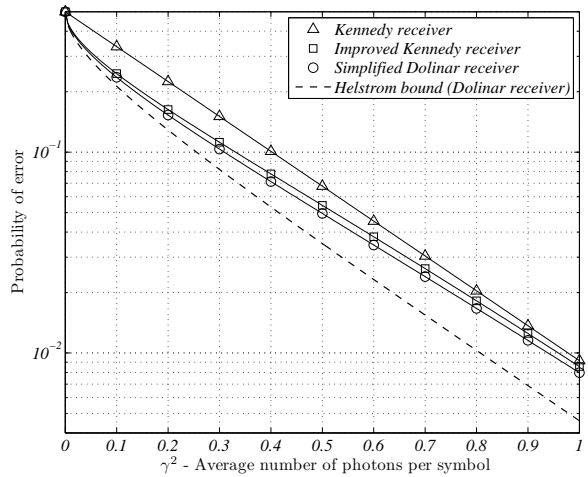}
    \caption{Symbol error probability in the case of equiprobable states ($q_0=q_1=1/2$).  The performance of the considered detection strategies are reported and compared with the Helstrom bound, which is achievable by the optimal Dolinar receiver. Emphasis is given to the quantum limited region corresponding to very weak coherent states.}
    \label{Fig:1}
\end{figure}

In Fig.~\ref{Fig:1} the performance of the Kennedy receiver and
of the improved Kennedy receiver are compared with the
Helstrom bound. (The  figure also includes the simplified Dolinar receiver that will be introduced in Section~\ref{Sect:5}). For large values of $\gamma$ the
improvement obtained by optimizing the displacement $\beta$
is negligible. On the other hand, as  $\gamma$ goes to 0,
both the Helstrom bound and $P_{IK}$ approach similar values.
Indeed, as shown in the figure,
the performance of the improved Kennedy receiver strictly
approximates the Helstrom bound for weak coherent states.

\section{A Simple Approach to the Dolinar Receiver}\label{Sect:4}
The above considerations suggest that, for $n$ large enough, such that copies of weak enough coherent states are obtained, the optimum measurements on the segmented states \eqref{eq:2} can be well approximated by suitable displacements and photon counting.
In other words, the sequences of measurement angles \eqref{eq:11} may
be reinterpreted as sequences of displacements. Then, the transition
to the continuous time--scheme depicted in Fig.~\ref{Fig:2}, appears to be natural.

The input field $\psi(t)$, $0<t<T$, corresponding to
the coherent state ${\ket(\pm\gamma)}$, is represented by
\begin{equation}
\psi(t)=\pm\psi e^{i2\pi f_0t}\;,\label{eq:17}
\end{equation}
where $f_0$ is the optical frequency and $T$ is the pulse
duration. The mean number of photons arriving to the detector
is given by
\begin{equation}
\gamma^2=\int_0^T|\psi(t)|^2dt=\psi^2 T\;.\label{eq:18}
\end{equation}
The detector subtracts from the input field a time--varying field
generated by a local laser with envelope chosen between either $u_0(t)$
or $u_1(t)$, accordingly to the value of $z(t)$, a binary
signal with possible values 0 and 1, giving the provisional
decision at time $t$. By mimicking the behavior
of the optimal \textit{multiple--copy} detection, we assume that the
decision signal $z(t)$ changes at any photon arrival at the
counter. Then, the optical signal at the  photon--counter
has envelope either $\pm \psi-u_0(t)$ or $\pm \psi-u_1(t)$, depending on the value of $z(t)$.
Moreover $z(T)$ is assumed to be the  final decision.

The mathematical problem is to choose the functions $u_0(t)$ and
$u_1(t)$ that maximize the correct detection probability
\begin{equation}
P_c=\P[z(T)=a]\;.\label{eq:20}
\end{equation}
The problem can be solved by means of standard photon counting statistics.

\newcommand*{\factor}{0.65}
 \begin{figure}
\psfrag{a}[][][\factor]{$\pm\psi$}
\psfrag{o}[][][\factor]{$\pm\psi-u_{z(t)}$}
\psfrag{dis}[][][\factor]{Displacement}
\psfrag{D}[][][\factor]{$D(-u_{z(t)})$}
\psfrag{ph}[][][\factor]{Photon}
\psfrag{cou}[][][\factor]{Detection}
\psfrag{z}[][][\factor]{$z(t)\in\{0,1\}$}
\psfrag{click}[][][\factor]{``click''}
\psfrag{Dec}[][][\factor]{Decision}
 \epsfxsize 1\hsize\epsfbox{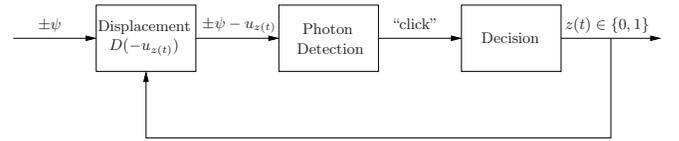}
\caption{Block diagram of the Dolinar receiver: The received signal with envelope $\pm \psi$ is displaced by a quantity $-u_{z(t)}$, where $z(t)\in\{0,1\}$ is the temporary estimation at time $t$. The value of $z(t)$ alternately changes from 0 to 1 at each single photon detection of the displaced signal $\pm \psi - u_{z(t)}$. The shape of the feedback signal $u_{z(t)}$ is also changed accordingly.}
    \label{Fig:2}
 \end{figure}
 
 %

Let us assume that $a=0$, so that $\psi(t)=\psi\, e^{i2\pi f_0t}$. Then, the
process $z(t)$ can be interpreted as a telegraph process \cite{Parzen}
alternately driven by non--homogeneous Poisson processes with
rates
\begin{equation}
\lb(t)=|\psi-u_0(t)|^2\qquad \text{and}\qquad \mu(t)=|\psi-u_1(t)|^2\;.\label{eq:21}
\end{equation}
A simple application of the properties of Poisson processes gives
the evolution of the conditional correct detection probability $p_0(t)=\P[z(t)=0|a=0]$. In fact, let $N(t,t+\Delta t)$ denote the number of photon arrivals at the counter in the interval $(t,t+\Delta t\,]$, therefore
\begin{equation}
\begin{split}
p_0(&t+\Delta t)= \\&=\Pc[z(t)=0,N(t,t+\Delta t)=0|a=0]\\&\hspace{8mm}+\Pc[z(t)=1,N(t,t+\Delta t)=1|a=0]
        +{\rm o}(\Delta t)\\
    &=\Pc[N(t,t+\Delta t)=0|z(t)=0]p_0(t) \\
     &\hspace{3mm} +\Pc[N(t,t+\Delta t)
        =1|z(t)=1][1-p_0(t)]+{\rm o}(\Delta t)\\
    &=[\,1-\lb(t)\Delta t\,]p_0(t)+\mu(t)\Delta t[1-p_0(t)]+{\rm o}(\Delta t)\;.
\end{split}\label{eq:22}
\end{equation}
Hence, the differential equation
\begin{equation}
p'_0(t)=\frac{\delta p_0(t)}{\delta t}=\mu(t)-[\lb(t)+\mu(t)]p_0(t)\;\label{eq:23}
\end{equation}
follows. In a similar way for $p_1(t)=\P[z(t)=1|a=1]$ we get
\begin{equation}
p'_1(t)=\frac{\delta p_1(t)}{\delta t}=\tilde\mu(t)-[\tilde\lb(t)+\tilde\mu(t)]p_1(t)
\label{eq:24}
\end{equation}
with
\begin{equation}
\tilde\lb(t)=|-\psi-u_1(t)|^2\qquad\text{and}\qquad \tilde\mu(t)=|-\psi-u_0(t)|^2\;.\label{eq:25}
\end{equation}

If our search is confined to symmetric solutions, namely, $u_1(t)
=-u_0(t)$ we get $\tilde\lb(t)=\lb(t)$ and $\tilde\mu(t)=\mu(t)$,
and the correct detection probability satisfies the differential
equation
\begin{equation}
\begin{split}
P_c'(t)&=\frac{\delta P_c(t)}{\delta t}=q_0p'_0(t)+q_1p_1'(t)\\&=\mu(t)-[\lb(t)+\mu(t)]P_c(t)\;.
\end{split}
\label{eq:26}
\end{equation}
On the basis of the results on \textit{multiple--copy} measurements
we expect that, for some choice $u(t)$ of the envelope of the feedback signal
$u_0(t)$, the provisional correct detection
probability $P_c(t)$ is exactly equal to the Helstrom bound applied
to the interval $(0,t)$, namely,
\begin{equation}
P_c(t)=\frac{1}{2}\left[1+\sqrt{1-4q_0q_1e^{-4\psi^2t}}\right]
\;.\label{eq:27}
\end{equation}
By substituting the above expression in \eqref{eq:26}, and defined $R(t)=\sqrt{1-4q_0q_1e^{-4\psi^2t}} $, we get
\begin{equation}
\begin{split}
&\psi^2\frac{1-R^2(t)}{R(t)}=\\&\hspace{5mm}=\psi^2+u^2(t)+2\psi u(t)-\left[\psi^2+u^2(t)\right]\left[1+R(t)\right]
\end{split}
\label{eq:29}
\end{equation}
and after some algebra
\begin{equation}
u(t)=\frac{\psi}{R(t)}=\frac{\psi}{\sqrt{1-4q_0q_1e^{-4\psi^2t}}}\label{eq:30}
\end{equation}
coinciding indeed with the Dolinar's solution (see also~\cite{Geremia}).


\section{A Suboptimal Receiver}\label{Sect:5}
The Dolinar receiver requires a time-varying feedback signal $u_0(t)$, whereas both the Kennedy receiver \cite{Kennedy} and its improved version \cite{Wittmann} make use of
a constant fixed displacement $\beta$ leading to a much simpler implementation.
Therefore, it is worthwhile to consider a simplified version of the Dolinar receiver where the feedback signal is constrained to have a constant fixed envelope $u_0(t)=\beta$. Such a setting would mean that only phase modulation, and specifically phase inversion, is required, whereas the optimal receiver has to also employ an amplitude modulator capable of generating an optical signal with shape defined by \eqref{eq:30}, which decays with $t$ but it is divergent about $t=0$ for the important case of equiprobable states $q_0=q_1=1/2\,$.

Hence, by substitution in \eqref{eq:25}, i.e., setting $\lambda(t)=\lambda=|\psi-\beta|^2$ and $\mu(t)=\mu=|\psi+\beta|^2$, with the initial condition $P_c(0)=q_0$, we get the following final probability of correct decision
\begin{equation}
P_c(T)
=\frac{1}{2}+\frac{\psi\beta}{\psi^2+\beta^2}+\left[q_0-\frac{1}{2}
-\frac{\psi\beta}{\psi^2+\beta^2}\right]e^{-2(\psi^2+\beta^2)T}\;.\label{eq:31}
\end{equation}
The optimized value of $\beta$ can be found by numerically solving the following transcendental equation, which is obtained by nulling the derivative of $P_c(T)$ made with respect to~$\beta$
\begin{equation}
\begin{split}
\beta T  (\psi^2 + \beta^2)  \left[ (2 q_0 -1)(\psi^2 +\beta^2) -2 \psi \beta \right] e^{-(\psi^2+\beta^2)T}=
\\=\psi(\psi^2 - \beta^2) \sinh((\psi^2 + \beta^2)T)\;.
\end{split}\label{eq:32}
\end{equation}

From Figure~\ref{Fig:1} we note that the simplified Dolinar receiver slightly outperforms the improved Kennedy receiver. Therefore, its experimental demonstration, and a study of its robustness to the presence of possible impairments, is an interesting task for future contributions.

\begin{figure}[t]
\centering
     \epsfxsize 0.9\hsize\epsfbox{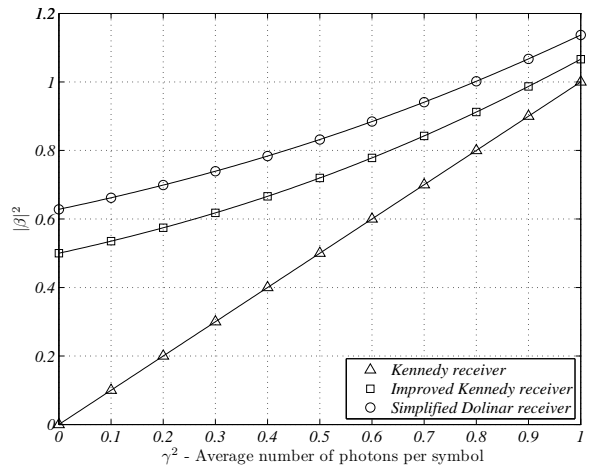}
    \caption{Intensity $|\beta|^2$ of the fixed displacement $D(-\beta)$ for the considered receivers; setting $q_0=q_1=1/2$, $T=1$.}
    \label{Fig:3}
\end{figure}

In Figure~\ref{Fig:3} the intensity of the displacements for different schemes is reported. It can be noted that, as confirmed by Figure~1, it is for very weak coherent signals that the simplified Dolinar receiver and the improved Kennedy receiver are particularly attractive, since, with the increasing of the signal strength, they both performs a displacement similar to the one applied by the original Kennedy's proposal, and both \eqref{eq:15} and \eqref{eq:31} approach~\eqref{eq:14}.

\section{Conclusions}\label{Sect:6}
A whole single coherent quantum state can be interpreted as a succession of many (possibly infinite) weaker copies of the same state, so that a different interpretation of the quantum discrimination task can be given. Such a view may provide further insights to well-established problems and give rise to novel detection solutions.

We have proposed an analysis of the behavior of the Dolinar receiver based on recent findings in the field of multiple-copy state discrimination. With such an approach it has been possible to provide a quite intuitive explanation, and simple mathematical derivation, of the Dolinar receiver. We also proposed a suboptimal simplified detection scheme that employs a photon counter, a phase inverter and, contrary to the Dolinar's solution, a constant-envelope displacement.

\begin{acknowledgments}
This work was supported in part by the Q-FUTURE project
(prot. STPD08ZXSJ), University of Padova.
\end{acknowledgments}

\newpage


\appendix*

\section{ Optimality of the multiple--copy adaptive measurement}

\medskip

The proof is done by induction. Using measurement vectors
\eqref{eq:7} with angle $\phi_1$, the correct detection
probability after the first measurement turns out
to be
\begin{equation}
\begin{split}
P_c^{(1)}&=q_0|\bk(\mu_{10},\gamma_0)|^2+q_1|\bk(\mu_{11},\gamma_1)|^2\\&=
q_0\cos^2(\theta-\phi_1)+q_1\sin^2(\theta+\phi_1)
\end{split}
\label{eq:A1}
\end{equation}
and, as it can be easily verified \cite{Acin}, it is maximized by
$\phi_1$ given by \eqref{eq:10} for $k=1$, with provisional
correct detection probability given by
\begin{equation}
P^{(1)}_c=\frac{1}{2}\left[1+\sqrt{1-4q_0q_1\chi^2}\right]\;.\label{eq:A2}
\end{equation}
Then, the result is proven for $k=1$. In particular,
simple computations give the a posteriori
probabilities $\Pc[a=i|z_1=i]=P_c^{(1)}$.

Now, suppose that the result holds true for $k-1$.
From the inductive hypothesis, the provisional
correct detection coincides with the Helstrom bound,
the adaptive measurement up to the $(k-1)$--th
copy coincides with the optimal global measurement
and the a posteriori probabilities are
\begin{equation}
\Pc[a=i|z_{k-1}=i]=P_c^{(k-1)}\;.\label{eq:A3}
\end{equation}
If these probabilities replace $q_0$ and $q_1$ in
the expression \eqref{eq:10} of the angle $\phi_k$ and in \eqref{eq:A2},
one gets
\begin{equation}
\begin{split}
P_c^{(k)}&=\frac{1}{2}\left[1+\sqrt{1-4P_c^{(k-1)}\left[1-P_c^{(k-1)}\right]\chi^2}\right]\\
&=\frac{1}{2}\left[1+\sqrt{1-4q_0q_1\chi^{2k}}\right]
\end{split}
\label{eq:A4}
\end{equation}
and the proof is complete.


\end{document}